# A Simple Method to Measure the Interaction Potential of Dielectric Grains in a Dusty Plasma

ZHUANHAO ZHANG, KE QIAO, JIE KONG, LORIN MATTHEWS, TRUELL HYDE

*Center for Astrophysics, Space Physics, & Engineering Research*

*Baylor University, Waco, Texas, 76706*

A simple minimally perturbative method is introduced which provides the ability to experimentally measure both the radial confining potential and the interaction potential between two individual dust particles, levitated in the sheath of a radio-frequency (RF) argon discharge. In this technique, a single dust particle is dropped into the plasma sheath to interact with a second individual dust particle already situated at the system's equilibrium point, without introducing any external perturbation. The resulting data is analyzed using a method employing a polynomial fit to the particle displacement(s), *X(t)*, to reduce uncertainty in calculation. Employing this technique, the horizontal confinement is shown to be parabolic over a wide range of pressures and displacements from the equilibrium point. The interaction potential is also measured and shown to be well-described by a screened Coulomb potential and to decrease with increasing pressure. Finally, the charge on the particle and the effective dust screening distance are calculated. It is shown for the first time experimentally that the charge on a particle in the sheath of an RF plasma decreases with increasing pressure, in agreement with theoretical predictions. The screening distance also decreases with increasing pressure as expected. This technique can be used for rapid determination of particle parameters in dusty plasma.

## I. Introduction

Dust particles immersed in plasma easily gain large negative charges due to the collection of electrons, which in general have higher velocities than do ions. In a GEC RF reference cell, dust particles come to equilibrium vertically above the lower powered electrode at the point where the gravitational and electric fields balance in the plasma sheath [1]. In the horizontal direction the dust particles interact with one another through interparticle electrostatic forces and are confined using an external potential; this results in the formation of crystalline structures [1, 2]. These structures have now been the subject of great interest for over a decade [3, 4, 5, 6, 7]. Particle-particle interactions are affected by the surrounding plasma, since this determines a particle's effective charge and the overall screening distance. Thus, a clear understanding of these interactions is crucial to any true understanding of the properties of the resulting crystals.

The interaction potential between dust particles is normally described by a screened Coulomb potential [8]

$$W(r) = \frac{Q^2}{4\pi\varepsilon_0} \frac{e^{-r/\lambda_{scr}}}{r} \qquad (1)$$

where $Q$ is the effective charge on the dust particle, $r$ is the inter-particle distance, and $\lambda_{scr}$ is the effective dust screening distance. Both the charge and screening distance can be calculated from a measurement of the interaction potential. However, numerical and experimental verification of this type of interaction usually requires a given set of assumptions (small particles, no plasma species flow, linear dispersion relation, etc.) and are restricted by experimental limitations (laser or probe perturbation, dust flow or clouds, etc.) as discussed below.

Konopka *et al.* [8, 9] analyzed the trajectories of two particles during head-on collisions. In this experiment, a bent Langmuir probe was introduced to manipulate the positions of the particles and introduce elastic collisions. The resulting interaction potential was then determined from the equations of motion. Melzer *et al.* [10] employed a laser manipulation method in order to study the interaction of two particles levitated simultaneously in the plasma sheath. The laser beam was focused on either the upper or the lower particle to induce relative motion; the resulting data showed there was a nonreciprocal interaction between the particles. Hebner *et al.* measured the interaction forces between particles constrained by a slot cut in a plate which was then placed on the bottom electrode [11, 12].

Hebner also utilized dust crystals as probes [13]. In each of these experiments, an external perturbation (probe, laser, slot, etc.) was introduced into the system in order to manipulate the particles, causing a disturbance of both the plasma and the particle-particle interaction potential. More recently, Fortov *et al.* [14] measured particle interaction with the help of gravity-driven probe particles, neglecting the interaction effect from surrounding particles.

The particle charge and screening distance also depend on the neutral gas pressure in the plasma chamber. Ratynskaia *et al.* [15] and Khrapak *et al.* [16] investigated computationally and experimentally the effect of elevated pressure on the charge of the particle in a DC bulk plasma employing very small grain particles (0.6 μm, 1.0 μm and 1.3 μm in radii). This work showed that the charge and screening distance both decrease with increasing pressure. The applicability of these results to particles levitated in the sheath of an RF plasma has been uncertain as there are significant differences between not only DC plasmas and RF plasmas, but also between the sheath region and the bulk plasma.

In this paper, a simple, minimally perturbative method is employed to simultaneously investigate both the radial confining potential and the particle-particle interaction potential. This is accomplished by introducing a single particle into the system which then collides with a second particle already residing at the equilibrium point within the sheath region. The damped motion of the resulting head-on collision is analyzed to determine the radial confining potential and the particle interaction potential. In contrast to previous work, this technique introduces no external perturbation of the system, allowing the plasma parameters to be reliably assumed as constant throughout the experimental period. This approach also allows the first experimental determination of the effect of pressure on the charge of the particle levitated in the sheath of an RF plasma under gravity.

## II Experimental setup

In this work, the confining potential and interparticle interaction potential for particles confined within a modified Gaseous Electronics Conference (GEC) RF reference cell [17] were examined. In the CASPER GEC cell (Fig. 1), a capacitively coupled discharge is produced between two parallel electrodes separated by 1.90 cm. The upper electrode consists of a grounded ring 8 cm in diameter, while a radio-frequency signal generator powers the lower electrode. The lower electrode has an aluminum plate with a milled circular cutout 1 mm deep and 25.4 mm in diameter to provide horizontal confinement for the dust. Melamine formaldehyde dust particles are used having a mass density of 1.514 g/cm$^3$ and a diameter of 8.89 ± 0.09 μm as provided by the manufacturer. These particles are illuminated by a vertical sheet of laser light and their trajectories recorded using a CCD camera running at 120 frames per second and having a resolution of 640 × 240 pixels. The plasma power was held constant at 1 W (13.56 MHz) throughout the experiment.

A dust dropper is used to drop single dust particles into the plasma sheath region where the falling dust particle collides with a second particle already located at the equilibrium point of the system. As part of the process of the newly formed two-particle system reaching an equilibrium state, the falling dust particle exhibits underdamped oscillations in the vertical direction and damped motion in the horizontal direction. Examination of the vertical oscillations of the falling particle allows calculation of the drag coefficient due to dust-neutral gas collisions employing a standard resonance frequency method [18] where, given the homogeneity of the system, it is assumed there is no significant difference between vertical and horizontal drag coefficients. The interaction between the two dust particles is measured directly, where the horizontal confining potential is quantitatively determined from the center of mass equations. Additionally, analysis of the direct particle-particle interaction allows determination of the interaction potential, providing both the charge and screening distance to be derived assuming a screened Coulomb potential structure. All of these are discussed in detail below.

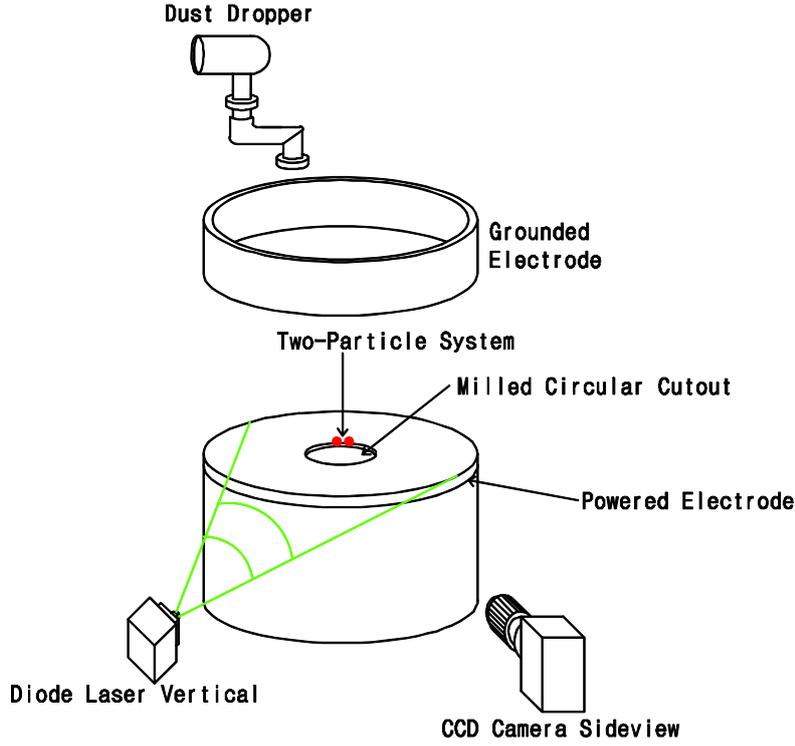

Fig. 1 Schematic diagram of experiments setup. Top electrode is grounded while the bottom one is powered, the separation between them is 1.90 cm. The interior radius of top electrode is 8 cm. The diameter of the milled circular cutout embedded in bottom electrode is 2.54 cm, and the depth is 0.10 cm. The particles are illuminated by a vertical sheet of laser and images are captured by a CCD camera from sideview.

## III Particle Motion in the Vertical Direction

As mentioned above, upon introduction into the plasma the falling dust particle initially undergoes damped oscillations in the vertical direction. These oscillations can be described as damped harmonic oscillations,

$$\ddot{z} + \beta \dot{z} + \omega_0^2 z = 0 \qquad (2)$$

where $\beta$ is the Epstein drag coefficient [19] due to the neutral drag force, $\omega_0$ is the natural frequency of the vertical potential well, and $z$ is the displacement of the particle's position from its equilibrium point. The amplitude of oscillation is given by the response function

$$R(\omega) = \frac{F_0}{\sqrt{(\omega_0^2 - \omega^2)^2 + \beta^2 \omega^2}} \qquad (3)$$

where $F_0$ is a constant and $\omega$ is the frequency of motion. Using the measured resonance curves, the Epstein drag coefficient can easily be determined by fitting the amplitude. Fig. 2 shows a representative damped oscillation with Fig. 3 showing the corresponding resonance curve and amplitude fit. As can be seen in Figure 4, the measured value for β increases linearly with increasing gas pressure, in good agreement with theoretical values obtained using the Epstein drag formula

$$\beta = \frac{8}{\pi} \frac{P}{\rho a v_{th,n}} \qquad (4)$$

where $P$ is the gas pressure, $\rho$ and $a$ are the mass density and radius of the dust particle, and $v_{th,n}$ is the thermal velocity of the neutral gas. At higher pressures (> 80 mTorr), vertical oscillations are already complete before the particle enters the camera frame; this results in the particle slowly approaching its equilibrium point from a horizontal direction without oscillation in either the vertical or horizontal direction. In this case, the drag coefficient is measured by first dropping a particle into the cell prior to ignition of the plasma. Once analysis of the particle's trajectory showed its velocity to be constant for a given pressure, the drag coefficient was calculated using the balance equation G = $m\beta v$, where $m$ and $v$ are the mass and velocity of the dust particle, respectively.

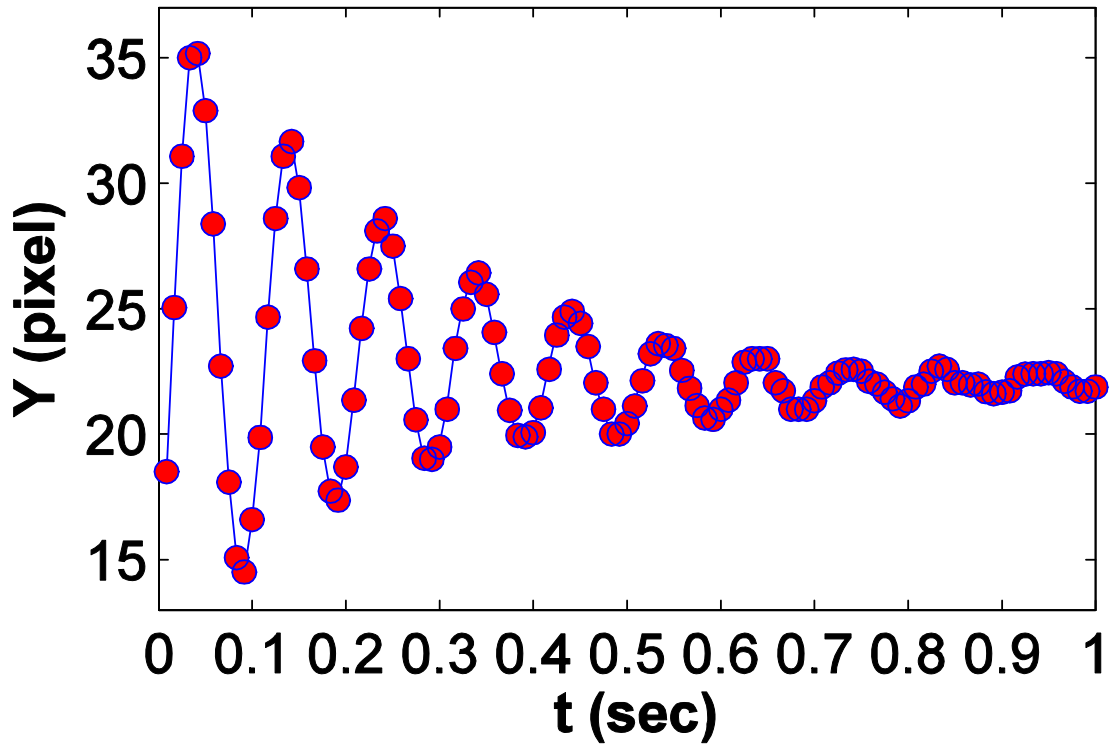

Fig. 2. Representative data showing damped oscillations in the vertical direction for a single dust particle introduced into the plasma as described in the text (P= 66 mTorr).

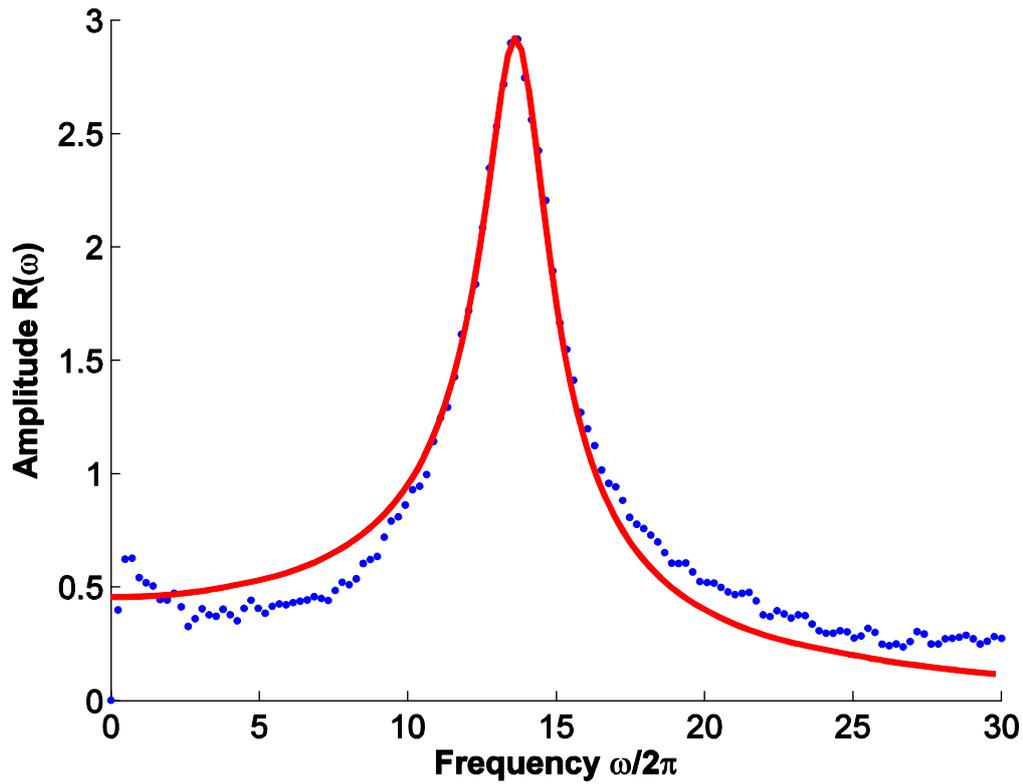

Fig. 3. Vertical resonance curves for the data shown in Figure 2. As shown, the measured resonance frequency is 13.2 Hz where the dots denote the direct Fourier Transform of the experimental data with the solid line showing a fit to the amplitude using equation (3). This fit provides a drag coefficient $\beta$ = 13.4 s$^{-1}$. (P = 66 mTorr.)

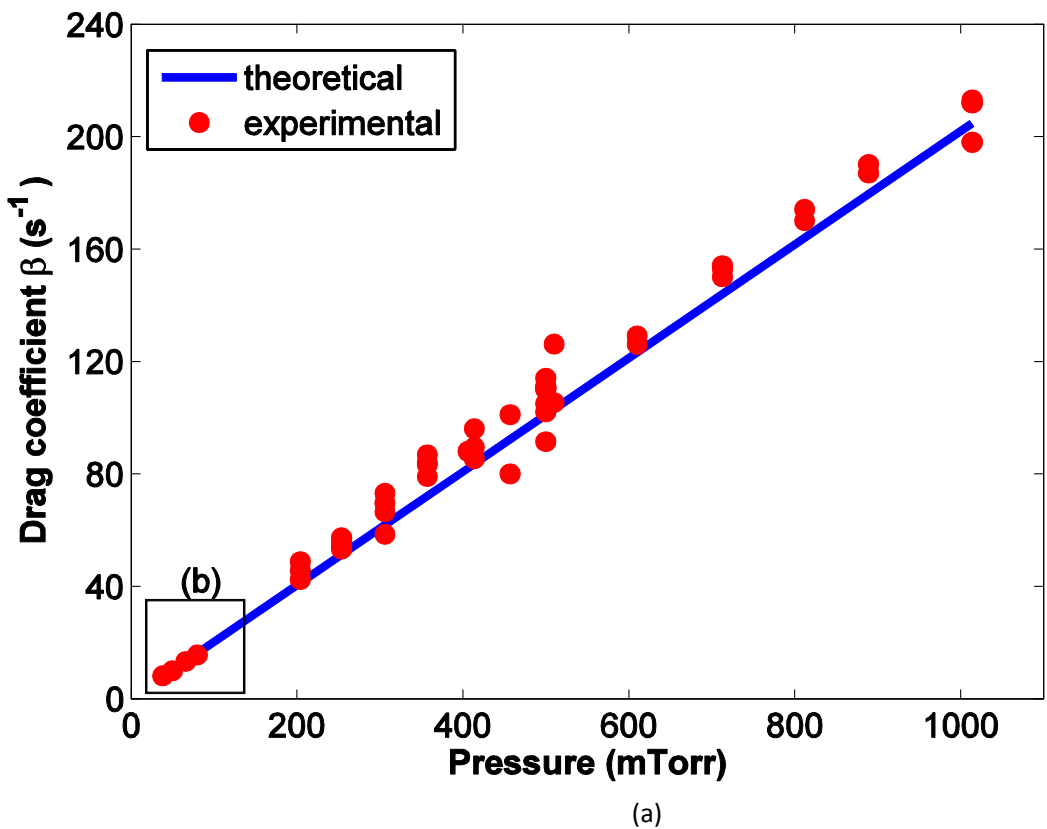

(a)

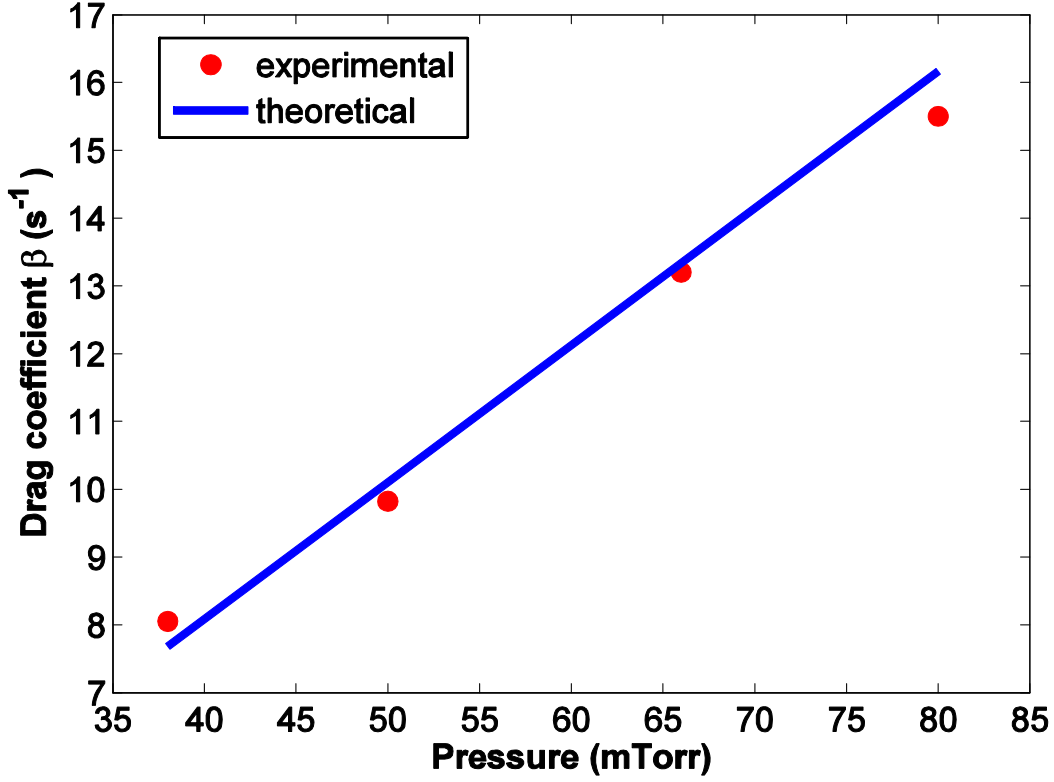

(b)

Fig. 4. Representative data showing the measured drag coefficient for various gas pressures for a dust particle in a plasma. Dots represent experimental data, while solid lines denote theoretical values as obtained from equation (4). The data acquired across all pressure ranges examined is shown in (a) while data collected for pressures below 85 mTorr is shown in expanded form in (b).

## IV Particle Motion in the Horizontal Direction

Particles within the CASPER GEC reference cell are confined horizontally by the horizontal component of the electric field created by the cutout in the plate placed on top of the powered bottom electrode. As mentioned earlier, a falling dust particle will undergo damped motion in the horizontal direction when interacting with the second particle via a screened Coulomb potential. Assuming the horizontal motion of the particle is along a straight line, and neglecting forces small when compared to the electrostatic forces involved (for example, the thermophoretic force or the radiation pressure force) the equation of motion [20] in the horizontal direction for the $i^{th}$ particle ($i$ = 1 or 2) is given by

$$\ddot{X}_i + \beta_i \dot{X}_i + \frac{1}{m_i}\frac{dW_{conf}}{dX}\bigg|_{X_i} = \frac{(-1)^{i+1}}{m_i}\frac{dW_{Inter}}{dX}\bigg|_{|X_r|} \quad (5)$$

where $X_i$ is the distance from the equilibrium point for particle $i$, $m_i$ is its mass, $\beta_i$ is the Epstein drag coefficient, $W_{conf}$ is the potential energy of the particle while in the confining region, $W_{Inter}$ is the interaction potential energy between the two particles, and the separation distance between the particles is given by $X_r = X_i - X_j$. Assuming identical particles, their drag coefficients and masses will be the same. Summing over particles 1 and 2 yields a center of mass, $X_C = (X_1 + X_2)/2$, with corresponding equation of motion

$$\ddot{X}_c + \beta \dot{X}_c + \frac{1}{m}\frac{dW_{conf}}{dX}\bigg|_{X_c} = 0. \quad (6)$$

The difference between these yields an equation for the relative motion of the two particles

$$\ddot{X}_r + \beta \dot{X}_r + \frac{2}{m}\frac{dW_{conf}}{dX}\bigg|_{X_r} = \frac{2}{m}\frac{dW_{Inter}}{dX}\bigg|_{|X_r|}. \quad (7)$$

where the confining potential is obtained through integration of equation (6)

$$W_{conf}(X_c(t_n)) = W_0 - \frac{m}{2}\dot{X}_c^{\,2}(t_n) - m\beta\int_{t_0}^{t_n}\dot{X}_c^{\,2}(t)dt. \quad (8)$$

Substituting the confining potential (8) into the equation of relative motion (7) and integrating, the particle interaction potential can be expressed as

$$W_{Inter}(X_r(t_n)) = W_{I0} - \frac{m}{4}\dot{X}_r(t_n)^2 - W_{conf}(X_r(t_n)) - \frac{m\beta}{2}\int_{t_0}^{t_n}\dot{X}_r^{\,2}(t)dt. \quad (9)$$

In the above, $W_0$ and $W_{I0}$ are both offsets. Defining the confining potential energy to be zero at the equilibrium point and assuming the interaction potential vanishes at infinite separation distance, $W_0$ and $W_{I0}$ can be determined, yielding both the confining potential and the interaction potential.

In (8) and (9), the majority of the uncertainty is associated with the measurement of the velocity (the first time derivative of position) and its associated time integral. To reduce this uncertainty, the time step was minimized using a camera frame rate of 120 Hz while the uncertainty in the velocity was minimized by fitting the center of mass equation (Fig. 5) and the relative motion equation (Fig. 6) by polynomials in $t$

$$X(t) = a_0 + a_1 t + a_2 t^2 + a_3 t^3 + a_4 t^4 + \ldots \quad . \quad (10)$$

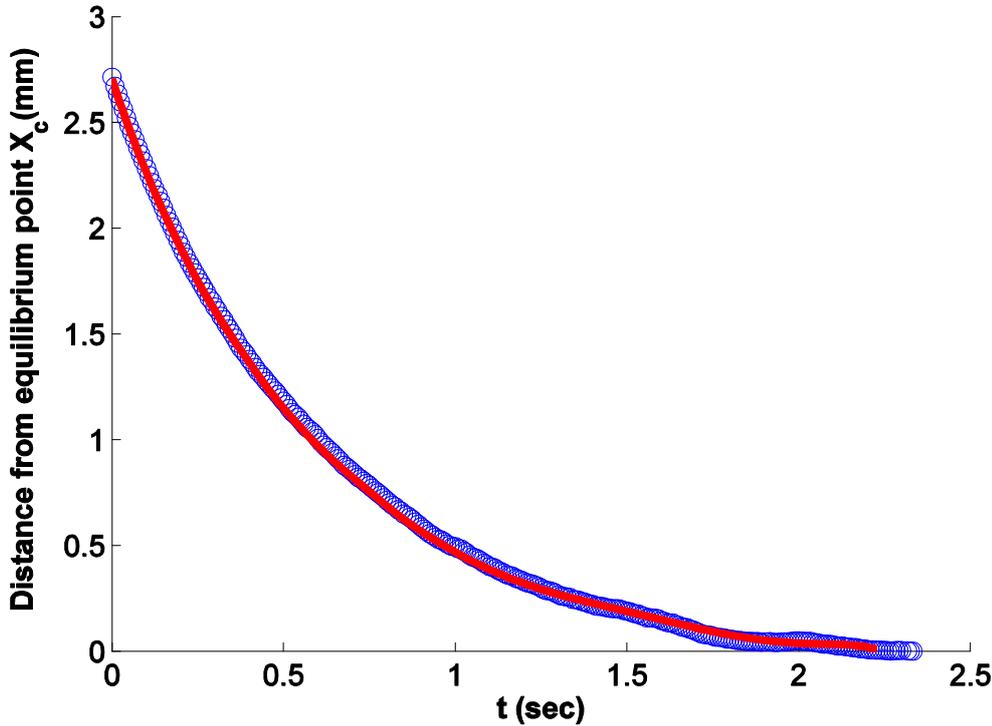

Fig. 5 (Color online) Center of mass distance from the equilibrium point as a function of time. The circles (blue) denote experimental data, while the solid line (red) shows a polynomial fit to $X_c(t)$. (P = 66 mTorr)

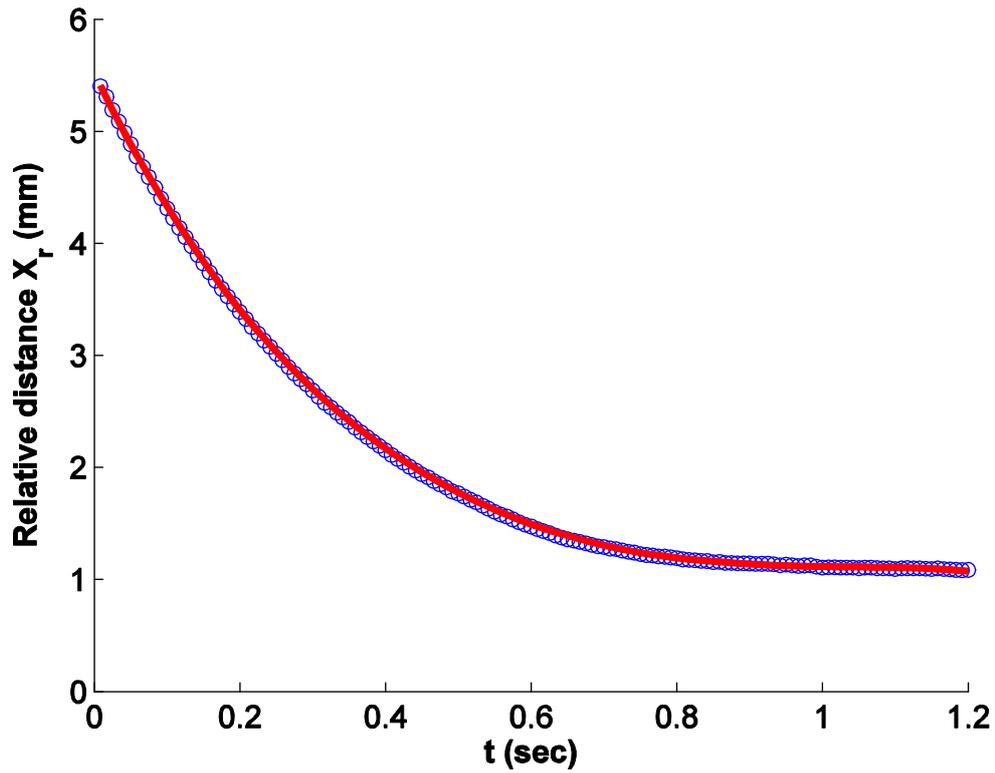

Fig. 6. (Color online) Relative distance between particles as a function of time. Circles (blue) denote experimental data, while the sold line (blue) shows a polynomial fit to $X_r(t)$. (P = 66 mTorr)

Fig. 7 shows the calculated horizontal potential energy corresponding to the data shown above. As can be seen, the potential is well-described by a polynomial fit using a parabolic function ($W_{conf}(X) = kX^2$) and these results are in good agreement with experimental data. Once the confining potential is determined, the interaction potential between the two particles can be easily found using equation (9) (Fig. 8). As seen, this interaction potential is well described by a screened Coulomb potential as expected. The particle charge and screening distance calculated from the experimental data and the polynomial fit are in good agreement, within the uncertainties described previously.

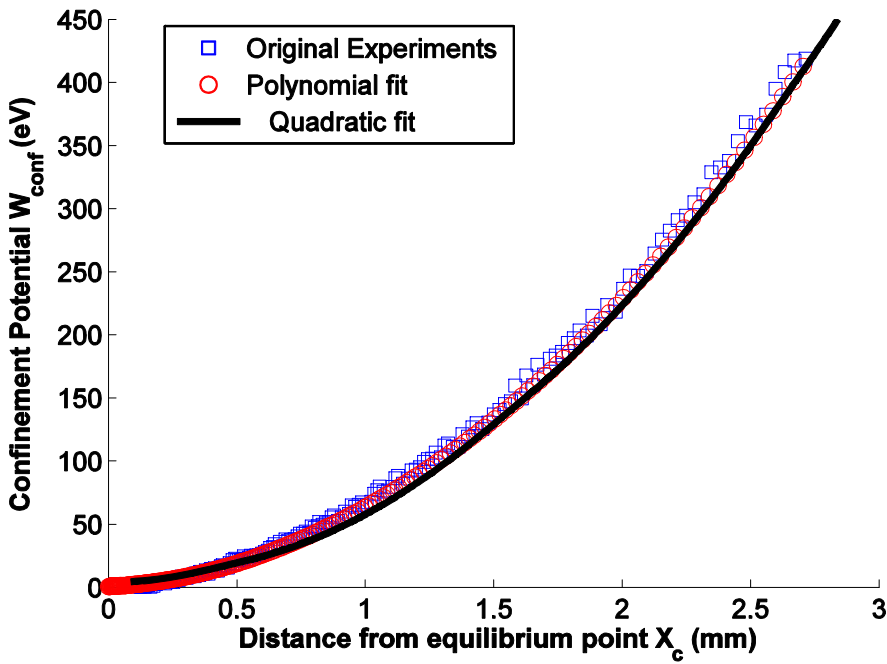

Fig. 7. (Color online) Measured horizontal potential energy. Squares (blue) denote values calculated from experimental data, while the circles (red) represent values calculated using a polynomial fit for $X_c(t)$. The solid (black) line shows a quadratic fit to the potential energy $W_{conf} = [52(X + 0.1)^2 - 0.4]$ eV where the resulting spring constant is $k = (8.33 \pm 0.18) \times 10^{-12}$ kg/s$^2$.

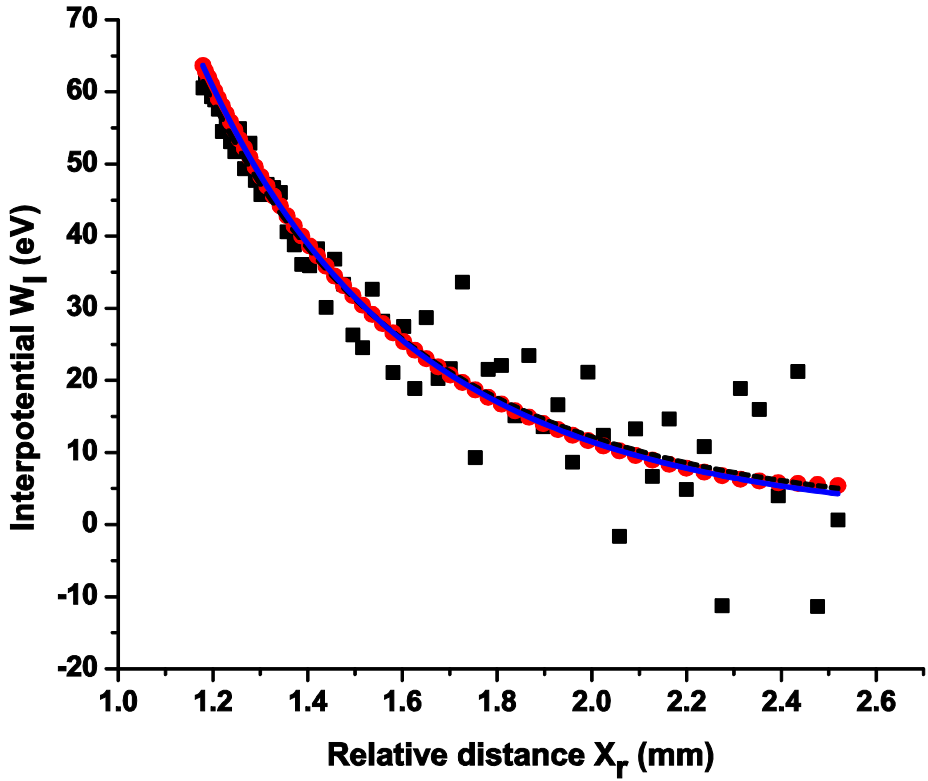

Fig. 8. (Color online) Measured interaction potential energy as a function of particle separation. Squares (black) represent values derived from the original experimental data while the dash line (black) shows the corresponding fit assuming a screened Coulomb potential. Using this fit, the calculated particle charge is $Q = (1.59 \pm 0.31) \times 10^4 e$ and calculated screening distance is $\lambda_{scr} = (729 \pm 192)$ $\mu m$. The dots (red) denote results derived using a polynomial fit $X_r(t)$. The solid line (blue) shows the corresponding screened Coulomb potential fit, with $Q = (1.69 \pm 0.06) \times 10^4 e$ and $\lambda_{scr} = (692 \pm 17)$ $\mu m$. (P = 66 mTorr)

## V Interaction Potential Relationship to Neutral Gas Pressure

The relationship between the interaction potential and the neutral gas pressure was also examined. As shown in Fig. 4, as the neutral gas pressure increases the drag force becomes stronger. This results in the center of mass motion changing from underdamped motion (P ≤ 38 mTorr) to critically damped motion (P ≈ 50 mTorr) and finally to overdamped motion (P ≥ 66 mTorr). The corresponding spring constants for the potential well are shown in Fig. 9. As shown, the spring constant is large across the under-damped regime, decreases sharply as the pressure increases to a critical point (P ≈ 50 mTorr) where the motion becomes critically damped, and then increases to a maximum ( ≈ 8.88 x 10$^{-12}$ kg/s$^2$ ) at a pressure of 80 mTorr. Beyond 80 mTorr, $k$ decreases in a uniform manner. Once the confining potential is determined, the interaction potential between the two particles can be calculated using equation (9). The data shown in Fig. 10 denotes these results, employing polynomial fits to the measured data. As shown, the interaction potential decreases with increasing pressure. The resulting screening distance and particle charge calculated from the interaction potential at various pressures are shown in Fig. 11 and Fig. 12 respectively. As can be seen, both decrease with increasing pressure.

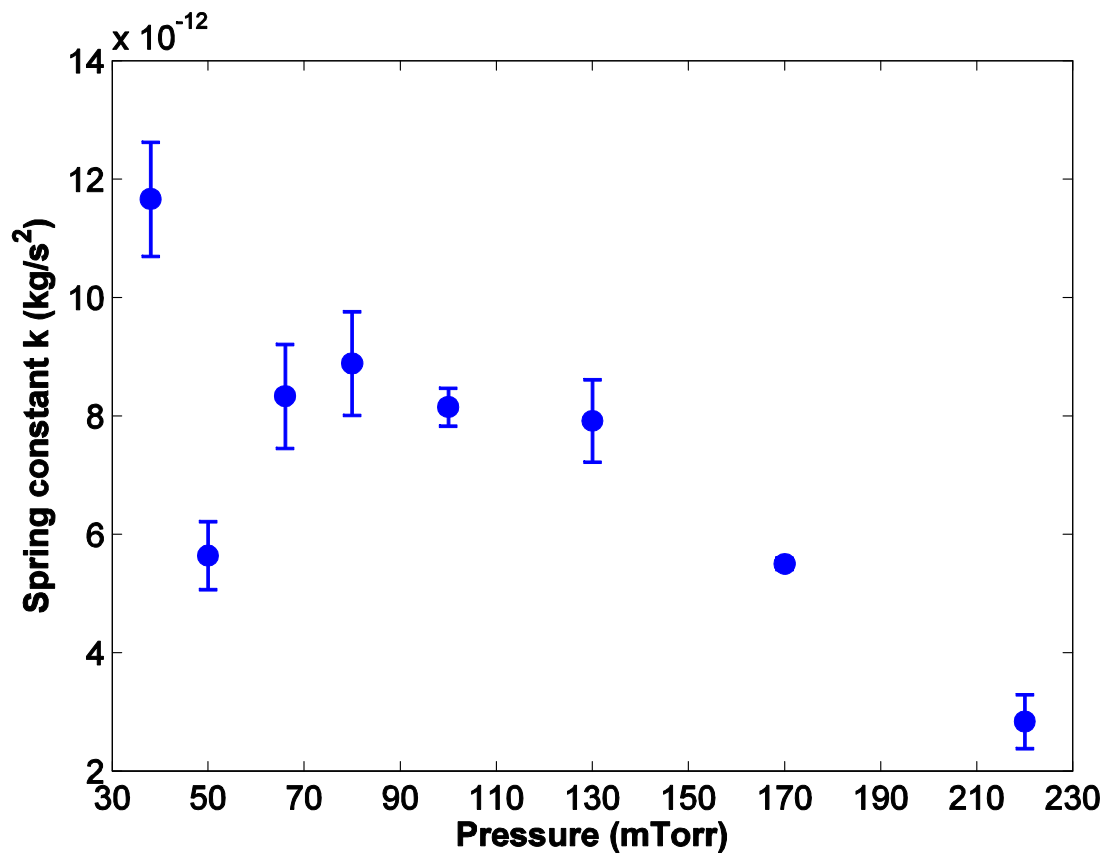

Fig. 9 Spring constants for the potential well at various pressures. A maximum value of k occurs at around 80 mTorr, with a minimum occurring between 38 mTorr and 50 mTorr.

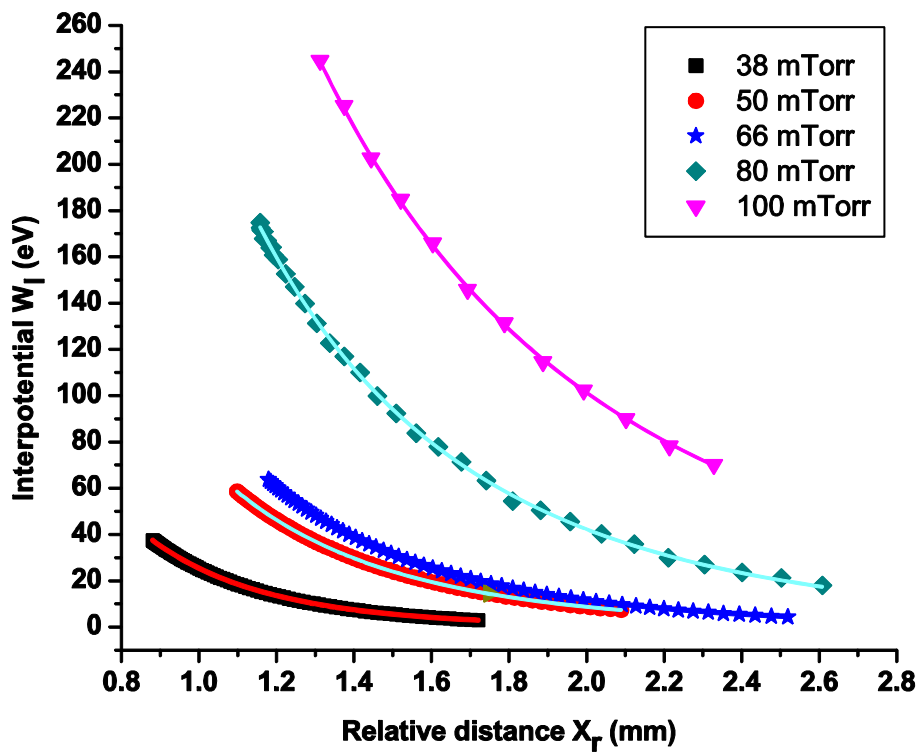

Fig. 10 (Color online) The interaction potential between two particles at various separations and pressures. Symbols represent results from a polynomial fit as described in the text, while solid lines show a screened Coulomb potential fit.

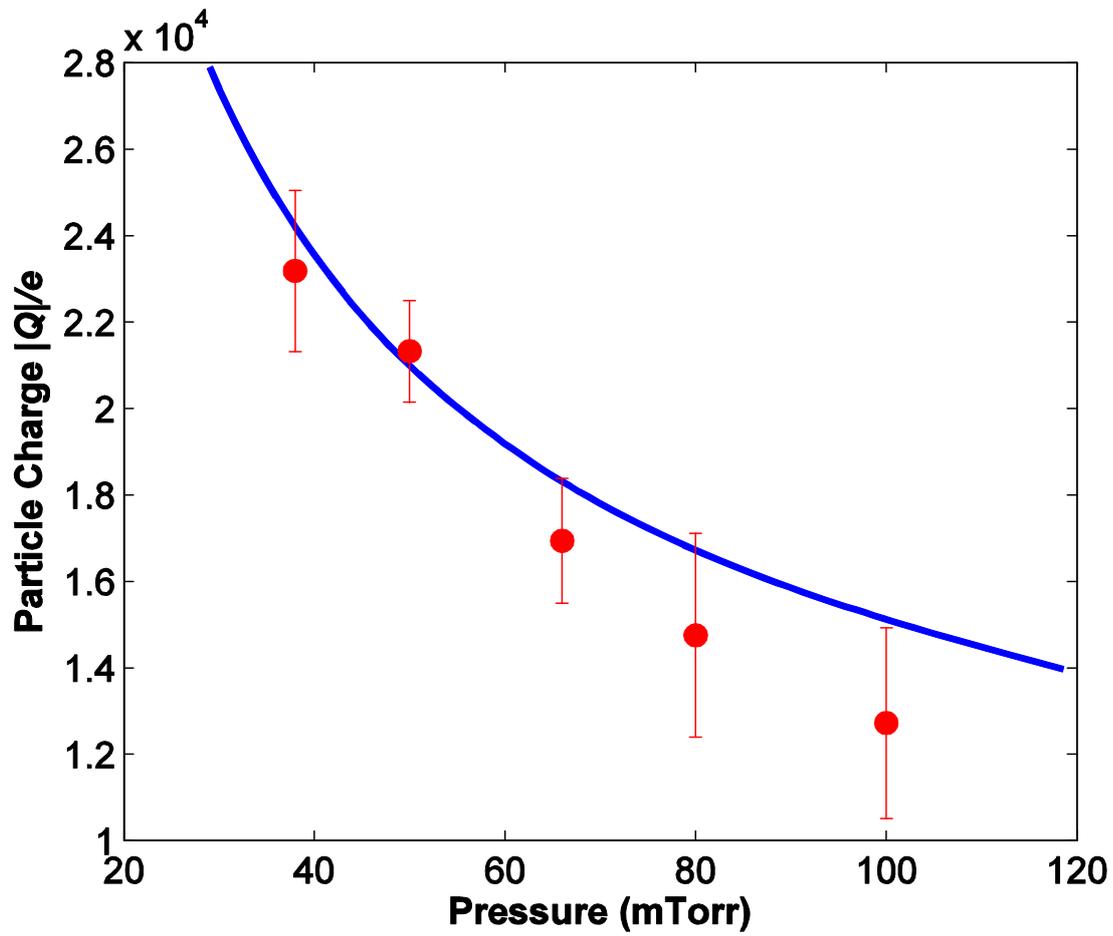

Fig. 11 Dust particle charge as a function of pressure. Dots denote experimental data while the solid line is generated theoretically as described in the text.

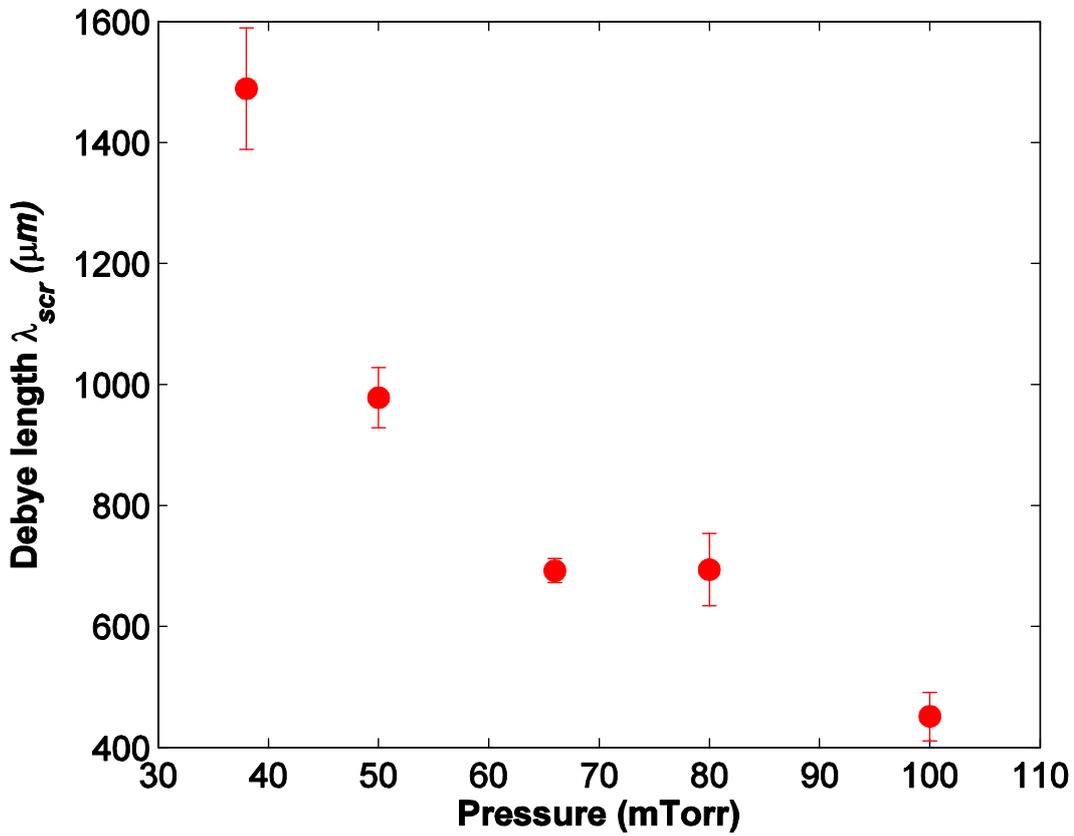

Fig. 12 Dust screening distance as a function of pressure. The dust screening length decreases with increasing pressure as explained in the text.

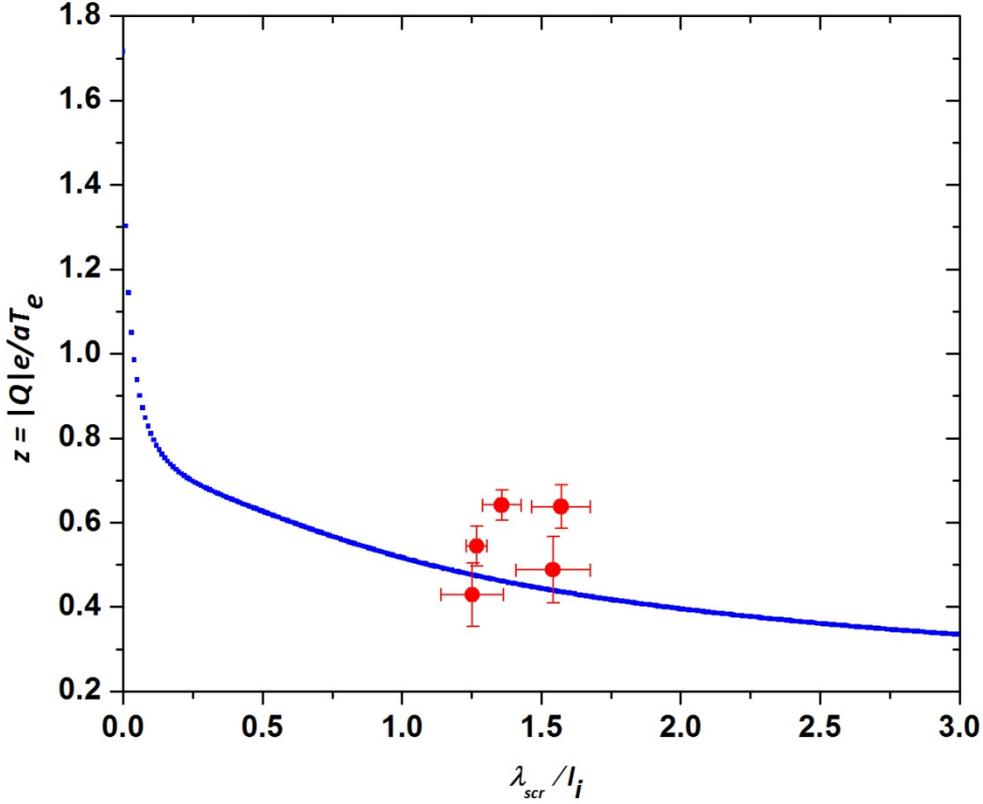

Fig. 13 (Color online) The normalized particle charge number $z = |Q|e/aT_e$ as a function of the ion collisionality index $\lambda_{scr}/l_i$. The smaller dots (blue) correspond to the theoretical calculation using Eqs. (11) and (12), while the bigger dots (red) come from these experimental results. The ion temperature, $T_i$, is assumed to be 0.03 eV while the electron temperature is, $T_e$ = 9.4848 + 14.68567*exp(-p/20.386) (eV) over the considered pressure range, where p is the gas pressure in unit of mTorr. The uncertainty of the probe measurements is not included here.

Assuming a weakly collisional regime, the net ion current to the dust particle surface can be expressed by [15, 16, 21]

$$I_i \approx \sqrt{8\pi}a^2 n_i v_{T_i}[1 + z\tau + 0.1z^2\tau^2(\lambda_{scr}/l_i)] \qquad (11)$$

where $a$ is the radius of dust particle, $n_i$ is the ion density, $v_{Ti}$ is the ion thermal velocity, $l_i$ is the ion mean free path, $\tau$ is the ratio of electron to ion temperature, $z$ is the dimensionless particle charge number in units of $e^2/aT_e$, and $\lambda_{scr}$ is the screening distance. This provides an electron current given by

$$I_e \approx \sqrt{8\pi}a^2 n_e v_{T_e} \exp(-z). \qquad (12)$$

where $n_e$ is the electron density and $v_{Te}$ is the electron thermal velocity. For an equilibrium state and assuming the ion and neutral gas temperatures are approximately $T_i \approx T_n \approx 0.03$ eV at room temperature, the particle charge can then be calculated by equating the ion and electron current $I_e=I_i$. These results are shown in Fig. 11. Fig. 13 shows the relationship between the normalized particle charge number and the ion collisionality index, both theoretically and experimentally.

The above results can be explained by considering the increased ion-neutral collision rate as the pressure increases. In Ref. [15], Khrapak *et al.* concluded that for a bulk DC discharge plasma, as this collision rate increases, the momentum of the ions is more easily transferred to the neutral gas resulting in slower moving ions. This results in more ions being captured by the negative dust grains. At the same time, the increasing pressure lowers the overall electron temperature resulting in a decreasing probability of electron-dust collisions. When combined, these reduce the overall negative charge on the dust. Additionally, the increased ion-neutral collision rate and lessened negative charge on the dust particle increases the local ion density in the vicinity of the dust particle, in turn increasing the shielding strength [22]. Thus, the screening distance is also reduced (Fig. 12). Where these trends have been predicted theoretically, the results are in good agreement with the experimental measurements shown in Fig. 11 and 13. The results in this paper are also close to recent data from Vaulina et al. [23, 24]. This is important since their technique considers inverse problem, i.e., describing the movement of dust particles employing Langevin equation, which also does not require outside perturbation. As far as the authors know, this is the first experimental verification of this effect in an RF plasma under gravity.

## VI  Conclusions

A simple method has been developed which allows simultaneous measurement of the confining potential well and the interaction potential for two particles in a complex (dusty) plasma. In contrast with previous work, no external perturbation (laser, probes, slot, etc.) to the system is required. This method also is advantageous in that it does not require any *a priori* assumptions for the analysis. For example, this method involves only a two-particle system; thus it doesn't rely on the assumption of a finite dust density and is not impacted by closely packed grains [13, 25]. Other experiments have used a wave method in particle flow to detect the particle charge, which requires an assumption of a linear dispersion relation [15, 26].

Employing the technique described in Section II, the confining potential well was confirmed to be parabolic (Fig 7, 9) and the interaction potential was shown to be a screened Coulomb potential over a wide range of pressures and separation distances from equilibrium (Figs. 8, 10). It was also shown that both the particle charge and the screening distance are dependent upon the neutral gas pressure, where both decrease with increasing pressure (Figs. 11 and 12) as predicted by theory. The increasing neutral gas pressure enhances the neutral-ion collision rate draining momentum from the ions; this results in ions with lower momentum(s) which are more easily captured by negatively charged dust grains. This has been shown experimentally within a DC plasma under microgravity conditions using the Plasma Kristall-4 (PK-4) facility on the international space station [15, 16, 21] but this is the first time it has been shown experimentally with in RF plasma under gravity. At the same time, the average electron temperature is lowered by increasing the pressure, leading to a lesser chance of electron-dust collisions. Both of these two effects reduce the net charge on the dust grains. At the same time, the density of ions are in the vicinity of the dust grains increases because of the smaller charge on the grains and increased ion-neutral collisions, resulting in the decrease of the screening distance [22]. In deriving equation (11), Khrapak [15] and Lampe [21] assumed no plasma species flow, i.e. dust grains immersed in the bulk plasma which can only be easily realized experimentally under microgravity. However, our experimental results show that the expression for ion current to the grain surface (Equation 11) can be extended to the case including ion flow in the sheath region under gravity conditions. Considering its simplicity and reliability, this method will be useful for a rapid determination of particle charge and screening distance, characteristics which play crucial roles in most dusty plasma experiments.

## Acknowledgments

The authors would like to thank Angela Douglass for data comparison and discussion, Victor Land for helpful suggestions, James Creel for probe measurements and Jorge Carmona Reyes for equipment setup.